\documentstyle[aps,twocolumn,]{revtex}
\input epsf

\begin{document}
\draft
\title{A magnetic tomography of a cavity state}
\author{R. Walser, J.I Cirac$^\dag$ and P. Zoller}
\address{Institut f{\"u}r Theoretische Physik,
Universit{\"a}t Innsbruck, 6020
Innsbruck, Austria\\
Departamento de Fisica Aplicada,
Universidad de Castilla--La Mancha, Ciudad Real, Spain$^\dag$,}
\date{August 22.,1996}
\maketitle{}

\begin{abstract}
A method to determine the state of a single quantized cavity mode is proposed. By adiabatic passage 
%%%\cite{parkins95}
the quantum state of the field is transfered completely
onto an internal Zeeman sub-manifold of an atom.
Utilizing a method of {\em Newton and Young} 
%%%%\cite{newton68} 
(Ann. Phys. {\bf 49}, 393 (1968)), we can determine
this angular momentum state uniquely, by a finite number of magnetic dipole
measurements with Stern-Gerlach analyzers.
An example illustrates the influence of dissipation.
\end{abstract}


\narrowtext
The state of a quantum mechanical system is completely specified by its
density operator ${\hat \rho}$. It is a fundamental as well as an important
practical question of quantum mechanics to devise measurement schemes which
allow a complete determination of ${\hat  \rho}$.
By a sequence of repeated measurements on an ensemble of 
identically prepared systems the state
has to be characterized operationally.
In quantum optics  this topic of complete state determination has received recently
considerable attention  in the context of characterizing nonclassical states of the
radiation field, and states of atomic and molecular motion 
\cite{vogel89,raymer93,lohmann93,walmsley95,wallentowitz95,schleich95,leonhardt95,buzek,wilkens95,cirac96}.

In a seminal paper, Vogel and Risken \cite{vogel89} have pointed out that the
state of a single mode of the radiation field (equivalent to a one-dimensional
harmonic oscillator) can be found by a tomographic techniques and
corresponding experiments have been performed by Raymer and
coworkers \cite{raymer93}. The central idea of quantum state tomography is
based on a reconstruction of the density matrix ${\hat \rho}$ from measured
quadrature probabilities $p_x(\theta)=(x,\theta|{\hat \rho}|x,\theta)$.
Here $|x,\theta)$ is a rotated eigenstate of the quadrature operator
${\bf x}_\theta$, i.e
$|x,\theta)={\cal R}(\theta)^\dagger\,|x\rangle=
e^{i\theta {\bf a}^\dagger {\bf a}}\,|x\rangle$ with $\bf a$ and 
${\bf a}^\dagger$
lowering and raising operators of the oscillator.
Alternative schemes have been discussed in the literature 
under the name of  state endoscopy \cite{schleich95} 
or by introducing discrete Wigner-functions \cite{leonhardt95}. 
Very recently, ideas were developed for a
state determination of ions moving in harmonic trapping potentials
\cite{wallentowitz95,cirac96}.

In the present paper we discuss a new scheme to measure the density matrix 
of the radiation field of a single quantized cavity mode by a 
magnetic tomography. It is based on combining ideas we have developed in the
context of quantum state engineering of arbitrary  Fock-state superpositions
in a cavity by adiabatic passage \cite{parkins95}, with a tomography of atomic
angular momentum states by Stern-Gerlach measurements originally proposed in \cite{newton68}.

It is well know that the adiabatic change of a 
Hamiltonian  interaction 
transforms initial energy-eigenstates into eigenstates of the final Hamilton
operator  \cite{messiah}.  This method to map quantum states is  applied in
various contexts of molecular- and atomic physics
 \cite{parkins95,bergmann89,marte91}.
It is also the key mechanism for transferring  the state of a quantized
cavity mode onto the internal state manifold of an atom.
%%%%%%%%%%%%%%%%%%%%%%%%%%%%%%%%%%%%%%%%%%%%%
\begin{figure}[ht]
\begin{center}
  \epsfxsize=6cm
    \hspace{0mm}
    \epsfbox{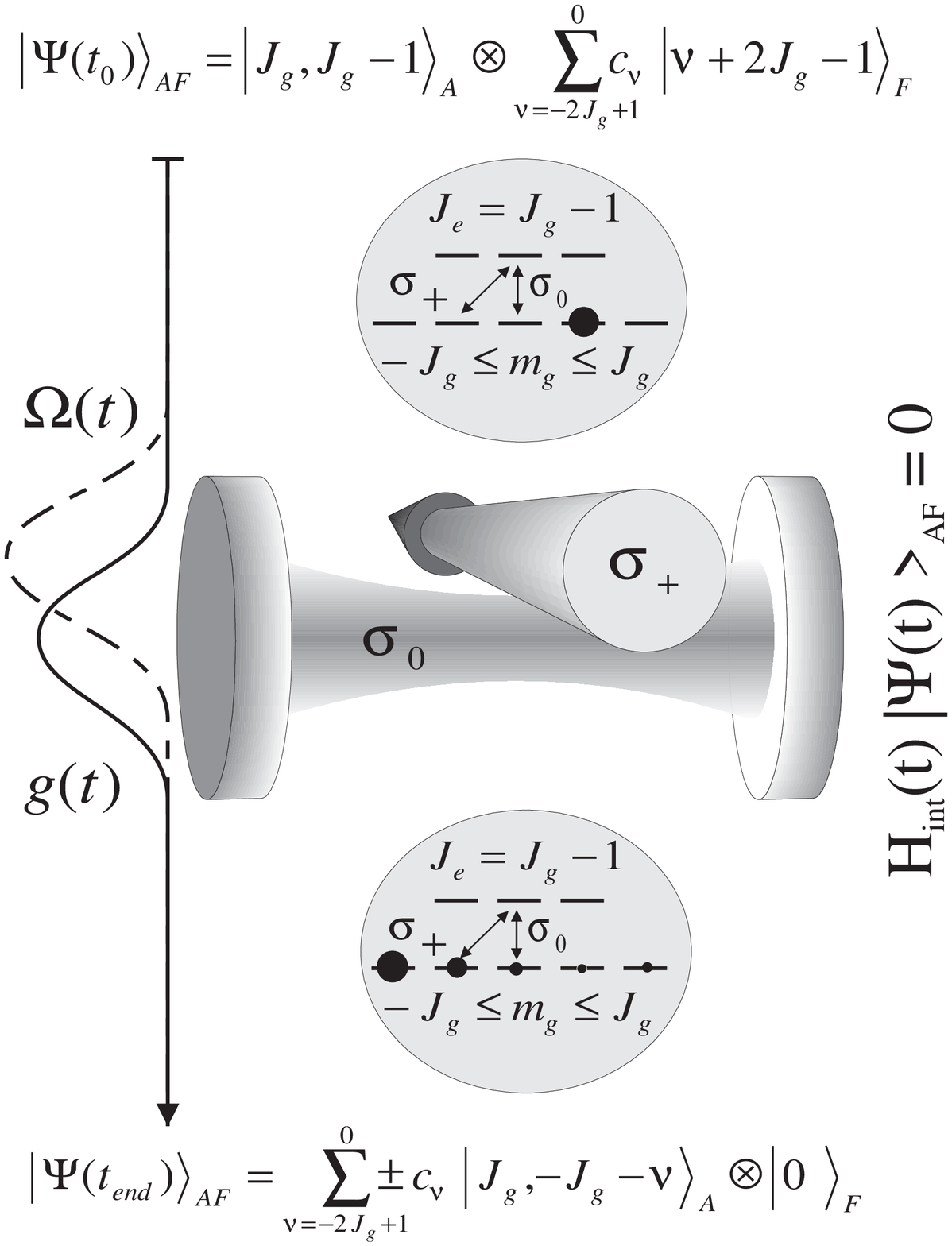}\\[0.3cm]
\begin{caption}
{\sf Evolution of
a degenerate two level atom (angular momentum: 
$J_g \rightarrow J_e=J_g-1$) coupled to a quantized 
$\sigma_0$-polarized cavity mode. First, it passes
through the profile a classical beam ($\sigma_+$) and then, with a delay,
through the cavity. }
                \label{figprep}
\end{caption}
\end{center}
\end{figure}
%%%%%%%%%%%%%%%%%%%%%%%%%%%%%%%%%%%%%%%%%%%%%%%%%%%%%%%%%%%%%%%%%%%%%%

According to Fig.~\ref{figprep},
an atom passes adiabatically \cite{parkins95,marte91} 
through the spatial profile
of a classical $\sigma_+$-polarized laser beam 
[Rabi-frequency: $\Omega(t)$] and, 
with a spatio-temporal displacement $\tau>0$ , through
the profile of a quantized, $\pi$-polarized cavity mode 
[atom-cavity coupling: $g(t-\tau)$].
We assume that the electronic structure of the atoms
corresponds to an optical $J_g\rightarrow J_e=J_g-1$ dipole transition.

The coupled atom-cavity system evolves according to the time-dependent Hamiltonian
\begin{eqnarray}
\label{hamadia}
&&H(t)/\hbar=\omega_c\, a_c^\dagger a_c+
\omega_{\rm eg}
\sum_{m_e=-J_e}^{J_e}|J_e,m_e \rangle_A \langle J_e,m_e|_A+\\\nonumber
&&\mbox{}
-i \Omega(t)(e^{i\omega_L t} A_1-A_1^\dagger e^{-i\omega_L t})
+i g(t-\tau)(a^\dagger A_0-A_0^\dagger a),
\end{eqnarray}
where $a_c$ and $\omega_c$ is the annihilation operator and oscillation 
frequency of the cavity mode, respectively. 
In terms of atomic basis states and Clebsch-Gordan coefficients
$C_{\sigma,m_g,m_e}^{1,J_g,J_e}$, the atomic de-excitation operators
$A_0, A_1$ 
are defined by
\begin{eqnarray*}
A_{\sigma}=\sum_{|m_g|\le J_g,|m_e|\le J_e}
|J_g,m_g \rangle_A \langle J_e,m_e|_A C_{\sigma,m_g,m_e}^{1,J_g,J_e}.
\end{eqnarray*}
The state space spanned by this Hamiltonian has the remarkable feature 
that it can be decomposed into invariant sub-spaces
${\cal H}=\bigoplus_{\nu=-2J_g}^{\infty}{\cal H}^{\nu}$. Due to angular
momentum conservation, it is only possible to couple
angular momentum states to a finite number of photon states by
means of a unitary evolution (Eq.~\ref{hamadia}). 
\begin{eqnarray*}
{\cal H}^{\nu}&=& 
\left\{|m_g|\le J_g,0\le n=\nu+J_g+m_g
\right\|\\
&&\left. |J_g,m_g\rangle_A\otimes |n\rangle_C,
|J_e,m_g\rangle_A\otimes |n-1\rangle_C\right\}.\nonumber
\end{eqnarray*}
One element of the sub-space ${\cal H}^\nu$ is of particular interest, i.e.
a linear combination involving only ground states \cite{bergmann89,vscpt}
\begin{equation} 
|\phi^\nu_0\rangle=\sum^{J_g-1}_{m_g=-J_g} \alpha_{m_g}^\nu |J_g,m_g\rangle_A
\otimes |\nu+J_g+m_g\rangle_C.
\end{equation}
By an appropriate choice of coefficients $\alpha_{m_g}^{\nu}$,i.e.
\begin{eqnarray}
\frac{\alpha_{m_g-1}^\nu(t)}{\alpha_{m_g}^\nu(t)}
&=&
\frac{g(t-\tau)}{\Omega(t)}
\sqrt{\nu+J_g+m_g}
\,\frac{C^{1,J_g,J_e}_{0,m_g,m_g}}{C^{1,J_g,J_e}_{1,m_g-1,m_g}},
\end{eqnarray}
this normalized state ($\alpha_{J_g-1}^{\nu}={\cal N}$) 
becomes also an eigenvector of the Hamilton operator
in the interaction representation (derived from Eq.~(\ref{hamadia})) 
and has a zero eigenvalue. 
According to the adiabatic theorem \cite{messiah}, this eigenvector approaches
a stationary eigenstate of the corresponding Schr\"odinger-equation, if
the time dependent change of the Hamilton operator during the total
interaction time $T$
is much less than the characteristic transition frequencies
($\Omega_{\rm max} T$, $\sqrt{N_{\rm max}}g_{\rm max} T\gg 1$).
Furthermore, if the delay and shape of the pulse sequences are chosen 
such that
\begin{equation}
0 \stackrel{-\infty\leftarrow t}{\longleftarrow}
g(t-\tau)/\Omega(t)
\stackrel{t\rightarrow +\infty}{\longrightarrow} \infty,
\end{equation}
then this is a mapping process that only permutes states
up to a sign change 
$s_\nu=\mbox{sign}(\alpha_{-J_g-\nu}^\nu(+\infty))$.
\begin{eqnarray*}
|J_g,J_g-1\rangle_A\otimes|\nu+2 J_g -1\rangle_C
\stackrel{-\infty \leftarrow t}{\longleftarrow} |\phi^\nu_0(t)\rangle,\\
|\phi^\nu_0(t)\rangle
\stackrel{t\rightarrow +\infty}{\longrightarrow}
s_\nu\, |J_g,-J_g-\nu\rangle_A \otimes |0\rangle_C.
\end{eqnarray*}

In other words, a coupled atom-cavity density operator
${\hat \rho}^{(AC)}$
that can be factorized initially into a pure atomic state
and a field state containing less than $2 J_g$ photons will 
be mapped to a product of atomic ground state superpositions 
and the cavity vacuum
\begin{eqnarray}
&&|J_g,J_g-1\rangle_A \langle J_g,J_g-1|_A\otimes {\hat \rho}^{(C)}
\stackrel{-\infty\leftarrow t}{\longleftarrow}
{\hat \rho}^{(AC)}(t)\\\nonumber
&&{\hat \rho}^{(AC)}(t)
\stackrel{t\rightarrow +\infty}{\longrightarrow}
{\hat \rho}^{(A)}
\otimes
|0\rangle_C \langle 0|_C,
\end{eqnarray}
with
\[
{\hat \rho}^{(C)}=\sum_{\nu,\mu=-(2J_g-1)}^{0} \rho_{\nu,\mu}^{(C)} \,
|\nu+2 J_g -1\rangle_C \langle  \mu+2 J_g -1|_C,\\\nonumber
\]
and
\[
{\hat \rho}^{(A)}=\sum_{\nu,\mu=-(2J_g-1)}^{0} s_\nu s_\mu\, 
\rho_{\nu,\mu}^{(C)} \,
|-J_g -\nu\rangle_A \langle  -J_g -\mu|_A.
\]
With reverse adiabatic passage, an internal atomic state is prepared
uniquely by reading out the cavity state.

\begin{figure}[ht]
\begin{center}
  \epsfxsize=5cm
    \hspace{0mm}
    \epsfbox{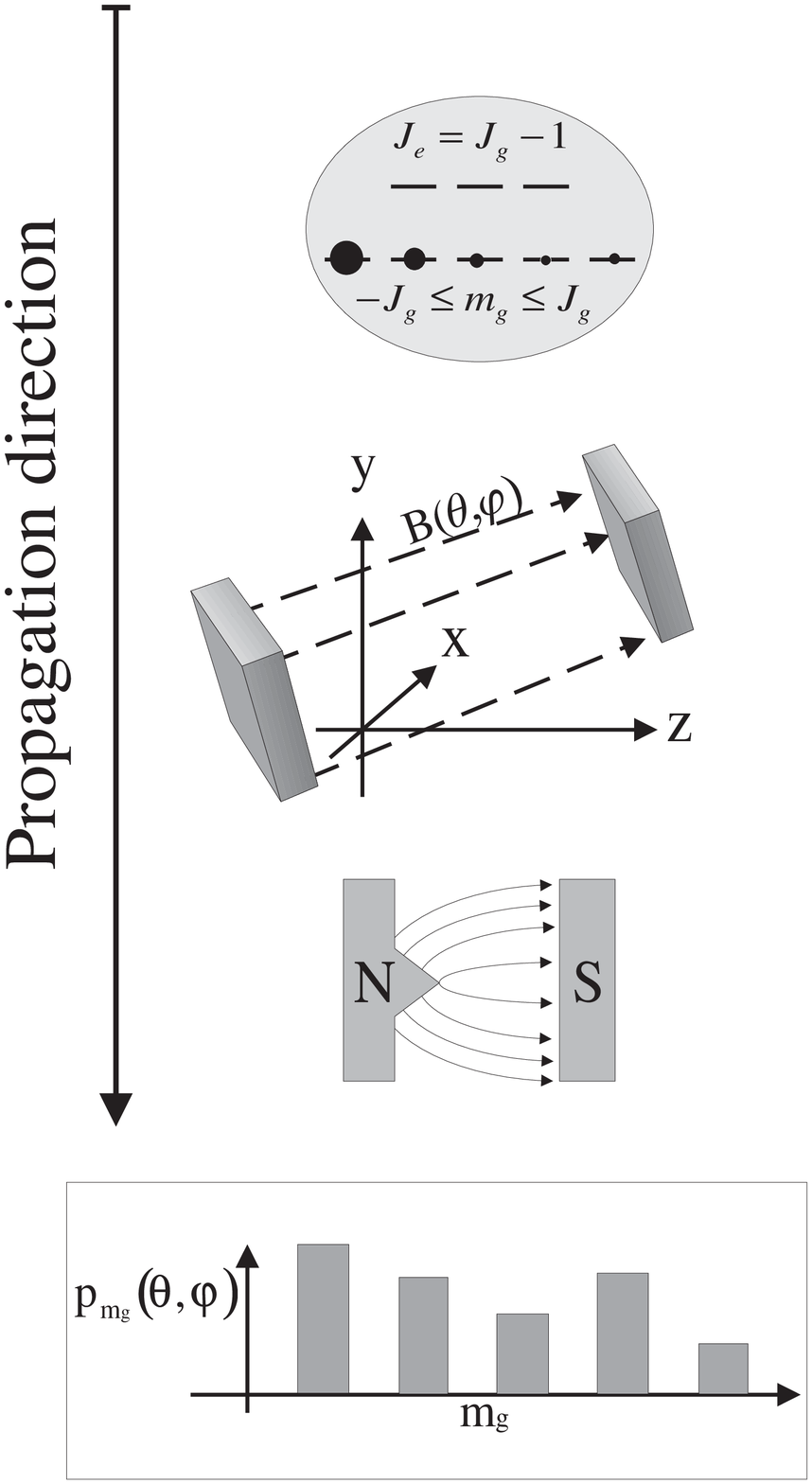}\\[0.3cm]
\begin{caption}
{{\sf Setup of a Stern-Gerlach experiment with an
additional homogeneous magnetic field 
${\vec B}(\theta,\varphi)=|B|\,{\vec n}(\theta,\varphi)$ inducing
spin precession around the axis ${\vec n}(\theta,\varphi)$}}
\label{setupsterger}
\end{caption}
\end{center}
\end{figure}

The complete characterization of such  an angular momentum state 
by a number of magnetic dipole measurements
was described in Ref.~\cite{newton68}.
It requires to  detect a set of physical observables that are proportional to
\[
\{ |m|\le J,|s| \le 2J\, \|{\hat P}_{m}(\theta,\varphi_s)\}.\] 
Here ${\hat P}_m(\theta,\varphi_s)$ represents the projector onto 
a rotated state 
$|J,m,{\cal R}(\theta,\varphi_s) {\vec e}_z\rangle
=D^{(J)}({\cal R})\,|J,m,{\vec e}_z\rangle$,
 $s$ enumerates an arbitrary set of 4J+1 azimuthal angles 
$\varphi_s$ and $\theta$ is a constant inclination.
This method can be implemented, for example, by
the unitary evolution of an angular momentum
state in a homogeneous magnetic field
${\vec B}(\theta,\varphi_s)=|B|\, {\vec n}(\theta,\varphi_s)$ 
oriented differently, each time the measurement is performed, and
by using a conventional Stern-Gerlach analyzer Fig.~\ref{setupsterger}.
%%%%%%%%%%%%%%%%%%%%%%%%%%%%%%%%%%%%%%%%%%%%%
%%%%%%%%%%%%%%%%%%%%%%%%%%%%%%%%%%%%%%%%%%%%%%%%%%%%%%%%%%%%%%%%%%%%%%

From the $J$-dimensional representation of the rotation group 
$D^{(J)}({\cal R})$ or the Wigner matrices $d^{(J)}(\theta)$
\cite{biedenharn}, one finds
\begin{eqnarray}
|J,m,{\cal R}(\theta,\varphi){\bf e}_z\rangle
&=&\sum_{n=-J}^{J} e^{-in\varphi}
d^{(J)}_{nm}(\theta) \,|J,n,{\bf e}_z\rangle.
\end{eqnarray} 
Hence, the occupation probabilities are  given by
\begin{equation}
\label{occpa2}
p_m(\theta,\varphi_s)=
\sum_{n,l=-J}^{J} e^{-i(n-l)\varphi_s} 
d^{(J)}_{nm}(\theta) d^{(J)}_{lm}(\theta)\,\rho_{ln}
\end{equation}
This linear equation relates density matrix
elements ($(2J+1)(2J+1)$ real numbers) to measured probabilities 
that are positive numbers. 
By determining $(2J+1)(4J+1)$ different probabilities,
this seemingly over determined set of linear equations 
has a unique solution that is positive definite. 
For $|l|\leq J$, $0\le w \le J-l$, one finds 
\begin{eqnarray}
\label{rhorec}
\rho_{l,l+w}=\sum_{m=-J}^{J}\sum_{j=0}^{2J}
C_{m, -m, 0 }^{J J j}
C_{l+w, -l, w}^{J J j}
\frac{(-1)^{l-m}}{d_{w, 0}^{(j)}(\theta)}
X(\theta)_{wm}.
\end{eqnarray}
In case of an equally spaced array of azimuthal angles $\varphi_s$, i.e,
$-\pi< \varphi_s=s \frac{2\pi}{4J+1}<\pi$, $|s|\le 2J$, 
the quantity $X(\theta)$ is the discrete Fourier transform of
the measured probability tableau
\begin{equation}
X(\theta)_{wm}=\frac{1}{4J+1}
\sum_{s=-2J}^{2J} e^{iw\varphi_s} p_m(\theta,\varphi_s ).
\end{equation}

In contrast to systems with continuous degrees of freedom,  
the reconstruction algorithm of Eq.~(\ref{rhorec})
is faithful if inclination angles $\theta$ are avoided where 
$d^{(J)}_{w,0}(\theta)$ vanishes (i.e. the zeros of an associated Legendre
polynomial). 
Most detrimental to this state tomography is the 
loss of cavity photons during the adiabatic interaction.
In contrast, spontaneous atomic decay is of minor importance as the adiabatic
eigenstate is formed by a ground state superposition.
To examine the influence of dissipation, we have 
coupled the atom-cavity system to an environment  \cite{parkins95}
and obtained the following master equation for the density
operator ${\hat \rho}$.
\begin{eqnarray}
\label{mastereq}
\frac{d}{dt}\rho=-\frac{i}{\hbar} 
[H_{\rm eff} \rho-\rho H_{\rm eff}^\dagger]  
+\Gamma \sum_{\sigma=0,\pm1} A_{\sigma} \rho A_{\sigma}^\dagger
+\kappa\, a_c\rho a_c^\dagger,
\end{eqnarray}
where $\Gamma$ and $\kappa$ denote the spontaneous decay rate 
and the inverse cavity life time, respectively.

In the interaction picture representation (derived from Eq.~\ref{hamadia}), 
the effective, non-hermitian Hamiltonian, introduced above, is given  by
\begin{eqnarray}
H_{\rm eff}&=&(\Delta -i\frac{\Gamma}{2})
\sum_{m_e=-J_e}^{J_e}|J_e,m_e \rangle \langle J_e,m_e|
-i\frac{\kappa}{2} \, a_c^\dagger a_c+\\\nonumber
&&\mbox{}
-i \Omega(t)(A_1-A_1^\dagger)+i g(t-\tau)(a_c^\dagger A_0-A_0^\dagger a_c).
\end{eqnarray}
For simplicity, it is assumed that the cavity 
and the external laser have a common frequency  $\omega_c=\omega_L$ and 
are detuned from the atomic resonance by 
$\Delta=\omega_{\rm eg}-\omega_L$.
The resulting atomic density operator can be determined either by solving 
the master-equation (Eq.~(\ref{mastereq}))  or alternatively, 
by averaging over a number of simulated quantum trajectories 
\cite{quantummontecarlo}. 
%%%%%%%%%%%%%%%%%%%%%%%%%%%%%%%%%%%%%%%%%%%%%
\begin{figure}[ht]
\begin{center}
  \epsfxsize=6cm
  \epsfysize=14cm
    \hspace{0mm}
    \epsfbox{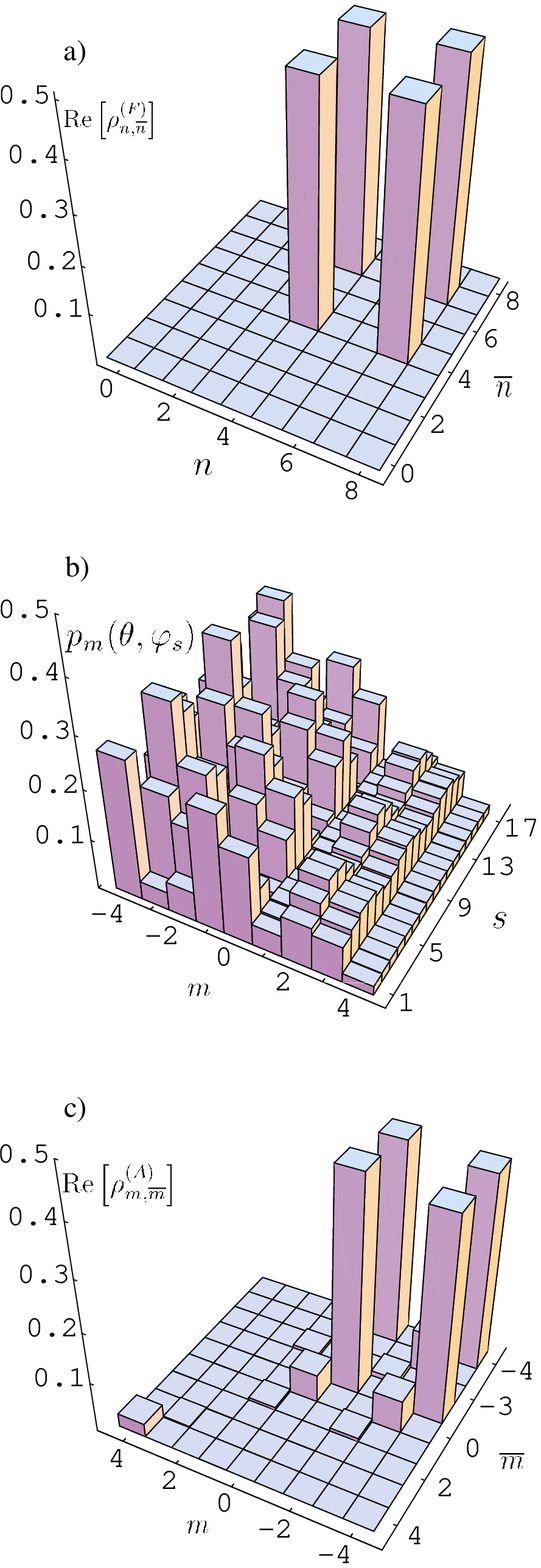}\\[0.2cm]
\begin{caption}
{\sf 
a) Real part of the initial cavity density matrix
$\rho_{n,{\overline n}}^{(F)}$ vs. photon number 
$n,{\overline n}$; b)~atomic ground state occupation probabilities 
$p_{m}(\theta,\varphi_s)$ vs. magnetic quantum number $m$ and
azimuthal phase index $s$ of an atom that moved adiabatically 
through the cavity;
c)~atomic density matrix $\rho_{m,{\overline m}}^{(A)}$
vs. magnetic quantum number $m,{\overline m}$, inverted 
tomographically from the data
represented in Fig.~\ref{figresults}b (see text for parameters).
}
\label{figresults}
\end{caption}
\end{center} 
\end{figure}
%%%%%%%%%%%%%%%%%%%%%%%%%%%%%%%%%%%%%%%%%%%%%%%

In Fig.~\ref{figresults}, the results of the mapping and reconstruction
process are shown for an initially pure cavity state
\begin{eqnarray}
|\psi(t_0)\rangle_C=\frac{1}{\sqrt{2}}(|4\rangle_C+|7\rangle_C), 
\end{eqnarray}
The real part of the initial cavity density matrix, i.e. 
$\rho^{(C)}(t_0)=|\psi(t_0)\rangle_C\langle\psi(t_0)|_C$
vs. photon number $n,{\overline n}$ is shown in Fig.~\ref{figresults}a.
In order to map this cavity state onto a Zeeman submanifold,
we assumed a sufficiently large degeneracy ($J_g=4\rightarrow J_e=3$).
Both fields are tuned to the atomic resonance  $\Delta=0$.
All frequencies are scaled to the spontaneous decay rate of 
the atomic excited state $\Gamma$, as the peak Rabi-frequency 
$\Omega_{\rm max}=50\, \Gamma$ and 
the maximal cavity coupling constant $g_{\rm max}=30\, \Gamma$. The time
dependent Gaussian turn-on (beam) profiles were of identical shape 
${\rm FWHM }=1\, [\Gamma^{-1}]$ and had a relative delay of 
$\tau=0.65\, [\Gamma^{-1}]$. 
To complete the adiabatic
passage, a total interaction interval
$(t_0=-1.07 \, [\Gamma^{-1}], t_{\rm end}=1.72 \,[\Gamma^{-1}])$ 
was chosen. The cavity decay
rate was set to $\kappa=0.01\, \Gamma$ which implies a decay probability of
$P_C=\kappa \langle N \rangle (t_{end}-t_0) \approx 0.15$ 
(for contemporary cavity QED experiments in the optical or microwave regime
see Ref.~\cite{walther90}).  

From the final density matrix, the atomic ground state 
occupation probability (Eq.~\ref{occpa2}) was evaluated with
$\varphi_s=s\frac{2\pi}{4J+1}$,  $|s|\leq2J$ and
\begin{eqnarray}
p_m(\frac{\pi}{3},\varphi_s)=
\langle J,m,{\bf n}(\frac{\pi}{3},\varphi_s)|\,
\rho_{\rm sim}^{(A)}(t_{\rm end})\,
|J,m,{\bf n}(\frac{\pi}{3},\varphi_s)\rangle.\nonumber
\end{eqnarray}
The result is depicted in Fig.~\ref{figresults}b.

Subsequently, we applied the tomographic inversion (defined in 
Eq.~\ref{rhorec}) to these data 
to obtain  $\rho_{m,{\overline m}}^{(A)}$, (Fig.~\ref{figresults}c).
Direct comparison with the simulated density matrix 
$\rho_{\rm sim}^{(A)}(t_{\rm end})$ shows that the inversion procedure
induces no error.
The additional features that appear in the mapped quantum state are of 
physical origin. The population $\rho_{4,4}^{(A)}$ is solely due to 
the spontaneous emission of a $\sigma_{-}$ photon, 
as this state is otherwise  not coupled to the dynamics at 
all. On the other hand, the satellite peaks that appear in vicinity of the
original coherent superposition state are caused by the decay of the
cavity state during the adiabatic mapping process. 
However, the resemblance with the original cavity state is striking.

In summary, we have studied a method to map the state of a single 
quantized cavity mode adiabatically onto a finite dimensional degenerate 
Zeeman submanifold of an atom that passes through the resonator.
Subsequently, we characterize this state by a number of repeated 
Stern-Gerlach measurements on identically prepared atoms as outlined 
in Ref.~\cite{newton68}. 
By a full quantum mechanical calculation, including spontaneous emission 
and cavity decay, we have shown that this method yields a faithfull
image of the original, a priori unknown cavity state. 
This method is not limited to the measurement of pure states
but may be applied also in case of statistical mixtures.

We would like to thank U. Leonhardt for stimulating discussions. 
R.W. acknowledges financial support from the Austrian FFW , 
Grant No. S6507-PHY.

%%% ------------------------ References ------------------------

\end{document}